\documentclass[aps,preprint,nofootinbib,floatfix,superscriptaddress]{revtex4}
\usepackage{epsfig}
\usepackage{graphicx}
\usepackage{multirow}
\usepackage{latexsym,amsbsy,amsmath}
\usepackage{stackrel}
\begin{document}
\title{$(I,J^P)=(1,1/2^+)$ $\Sigma NN$ quasibound state}
\author{H.~Garcilazo} 
\email{hgarcilazos@ipn.mx} 
\affiliation{Escuela Superior de F\' \i sica y Matem\'aticas, \\ 
Instituto Polit\'ecnico Nacional, Edificio 9, 
07738 Mexico D.F., Mexico} 
\author{A.~Valcarce} 
\email{valcarce@usal.es} 
\affiliation{Departamento de F\'\i sica Fundamental,\\ 
Universidad de Salamanca, E-37008 Salamanca, Spain}
\date{\today} 

\begin{abstract}
 JLab has recently found indications of the possible existence of a 
$\Sigma NN$ resonance at $(3.14 \pm 0.84) - i (2.28 \pm 1.2)$ MeV.
In the past, using models that exploit symmetries between the two-baryon sector
with and without strangeness, hyperon-nucleon interactions have been derived
that reproduce the experimental data of the strangeness $-1$ sector.
We make use of these interactions to review existing Faddeev 
studies of the $\Lambda NN -\Sigma NN$ 
system that show theoretical evidences about 
a $(I,J^P)=(1,1/2^+)$ $\Sigma NN$ quasibound state near threshold.
The calculated position of the pole is at 2.92$\, -i\, $2.17 MeV, 
in reasonable agreement with the experimental findings.
\end{abstract}
\maketitle

Hall A Collaboration at Jefferson Lab has made use of the 
$(e,e^\prime K^+)$ reaction to study the possible existence 
of neutral three-body $\Lambda$ and $\Sigma$ hypernuclei~\cite{Pan22}. 
They reported an excess of events around the $\Sigma$ thresholds.
The most significant enhancement appears $3.14 \pm 0.84$ MeV below 
the $\Sigma^0nn$ threshold and has a width of $\sigma \approx 2.28 \pm 1.2$ MeV. It
possibly hints at a bound $\Sigma^0 nn$ (I = 1) state. 

The existing experimental data and the expected 
forthcoming optimized data call for theoretical studies that 
could help with their interpretation.
In this letter it is our purpose to emphasize the relevant findings of 
existing Faddeev studies of the $\Lambda NN -\Sigma NN$ system. 
The theoretical results obtained are a valuable tool 
to analyze the Hall A Collaboration data.

We have carried out a detailed study of the $\Lambda NN-\Sigma NN$
three-body system at threshold looking for 
bound states or resonances~\cite{Gar07}. 
The strangeness $-1$ two-body interactions have been derived 
from the chiral quark cluster model (CQCM)~\cite{Val05}, by exploiting
the symmetries with the two-nucleon sector. 
In the CQCM hadrons are clusters of massive (constituent) quarks. 
As color carriers, massive quarks are confined through a confining potential.
They interact through a one-gluon exchange potential arising from 
Quantum Chromodynamics (QCD) perturbative effects. The non-perturbative effects
generate one-boson exchange potentials between quarks~\cite{Val05,Gac07}.

The nucleon-nucleon ($NN$) and hyperon-nucleon ($YN$) interactions
describe reasonably well the $NN$ and $YN$ two-body observables~\cite{Gac07}.
In particular, the two-nucleon system low-energy parameters, the $NN$ $S-$wave 
phase shifts, and the triton binding energy are described correctly~\cite{Val05}.
Besides, there is a reasonable agreement with the hyperon-nucleon 
elastic and inelastic scattering cross sections and the hypertriton 
binding energy. Finally, the isospin one $\Lambda nn$ system is 
unbound~\cite{Gar07,Gac07,Gar87}.

At the two-body level, the $N\Lambda-N\Sigma$ coupling as well as the 
tensor force, responsible for the coupling between $S$ and $D$ waves,
have been considered. The $\Lambda \leftrightarrow \Sigma$ conversion
is crucial to have a correct description of the $\Lambda NN$ system~\cite{Gar07}.
The $NN$ and $YN$ interactions contain sizable non-central terms which are
responsible, among others, for the deuteron binding energy.
The relevance of the $YN$ tensor force becomes apparent when studying
the $\Sigma^- p\to\Lambda n$ reaction. Such process is controlled
by the $\Sigma N(\ell=0)\to \Lambda N(\ell=2)$ transition
so that if only the central interaction 
$\Sigma N(\ell=0)\to \Lambda N(\ell=0)$ is considered, the cross section
cannot be described correctly~\cite{Str95}. 
The non-central $N\Lambda-N\Sigma$ interaction
induces a three-body force through the coupling between $YNN$ channels 
with $(\ell,\lambda)=(0,0)$ and $(\ell,\lambda)=(2,2)$, where the $YN$ 
relative orbital angular momentum is denoted by $\ell$ and $\lambda$ 
stands for that of the spectator nucleon respect to the $YN$ system.

For this study different models have been designed by choosing
sets of spin-singlet and spin-triplet $\Lambda N$ scattering 
lengths describing correctly the available experimental data.
In particular, besides a reasonable description of the $YN$ 
cross sections, the hypertriton binding energy corresponds to its
experimental value within the error bars 
$B_{0,1/2}=0.130\pm 0.050$ MeV~\cite{Jur73}.
The upper limit of the $\Lambda N$ spin-triplet scattering length,
$a^{\Lambda N}_{1/2,1}$, has been
established by requiring that the $(I,J^P)=(0,3/2^+)$ $\Lambda NN$ 
state does not become bound~\cite{Gar07}. 
The lower limit was set by requiring a correct description
of the $YN$ cross sections, which deteriorates markedly as
the $\Lambda N$ spin-triplet scattering length decreases.
Thus, it was found that $1.41 \le  a^{\Lambda N}_{1/2,1} \le 1.58$ fm. 
Once the $\Lambda N$ spin-triplet scattering length has been defined, 
the $\Lambda N$ spin-singlet scattering length,
$a^{\Lambda N}_{1/2,0}$, was constrained by
demanding that the hypertriton binding energy is in the experimental interval $B=0.130\pm 0.050$
MeV, leading to $2.33 \le a^{\Lambda N}_{1/2,0} \le 2.48$ fm.  
Without loss of generality we take the model
with $a^{\Lambda N}_{1/2,1}=1.41$ and $a^{\Lambda N}_{1/2,0}=2.48$ 
as the reference model. All calculations have been performed for several
models within the scattering lengths intervals and the conclusions remain 
unchanged. For the reference model the hypertriton binding energy
obtained is 129 keV. Just to illustrate the relevant role played by the
$D$ waves of the three-body system, it is worth to note that considering 
only $S$ wave three-body channels the hypertriton binding energy is 89 keV,
out of the experimental range.

The solution of the three-body problem have been 
described elsewhere~\cite{Gar07} and are out of the scope of this letter.
We focus on the results concerning the possible existence of a $\Sigma NN$ $(I,J^P)=(1,1/2^+)$
resonance~\cite{Pan22}.
\begin{figure}[t]
\vspace*{-0.5cm}
\includegraphics[width=.58\columnwidth]{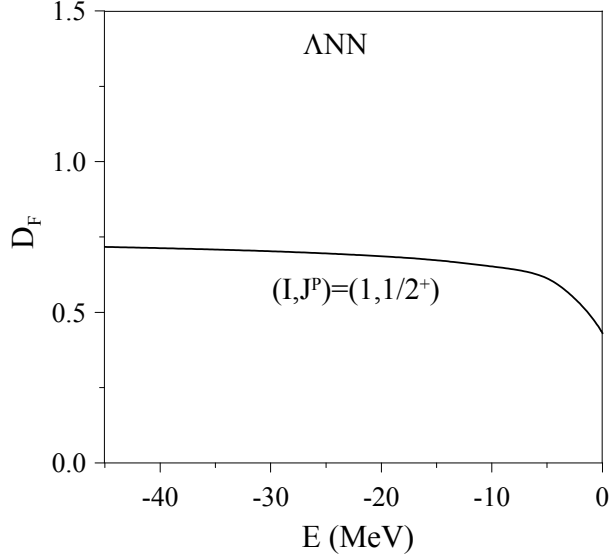}
\vspace*{-6.cm}
\caption{Fredholm determinant, $\rm{D_F}$, for the $J^P=1/2^+$ $\Sigma NN$
channels utilizing the reference model, in which the
deuteron binding energy is $E=-2.225$ MeV~\cite{Gar07}.}
\label{fig4}
\end{figure}

Let us first discuss the attractive or repulsive character of the 
different $J^P=1/2^+$ $\Sigma NN$ channels.
Fig.~\ref{fig4} shows the Fredholm determinant of the 
$J^P=1/2^+$ $\Sigma NN$ channels below the 
$\Sigma d$ threshold, where the continuum starts. 
The Fredholm determinant of the $I=0$ and $1$ channels is
complex because the $\Lambda NN$ channels are open.
The imaginary part is small and uninteresting. 
It can be seen that the channel showing the most attractive character
is the $(I,J^P)=(1,1/2^+)$. For an 
attractive channel the Fredholm determinant, $D_F$, is smaller
than 1 and it becomes negative if a
bound state exists~\cite{Gar87}. Thus,
as the Fredholm determinant is very close to zero
at the $\Sigma d$ threshold is a clear indication of 
a quasibound state. The $(I,J^P)=(0,1/2^+)$ channel is also 
attractive, but far less than the $I=1$ one.
This can be easily understood as follows.
We show in Table~\ref{t4} the two-body channels that contribute to
a given $J^P=1/2^+$ $\Lambda NN - \Sigma NN$ state with isospin $I$.
The most attractive two-body channels, in particular
the $\Sigma N$ $^{3}S_{1}(I=1/2)$ and 
$^{1}S_{0}(I=3/2)$ and the $NN$ $^{3}S_{1}(I=0)$,
contribute to the $(I,J^P)=(1,1/2^+)$ $\Sigma NN$ state,
however, the last two are forbidden for the $(I,J^P)=(0,1/2^+)$ $\Sigma NN$
state, one of them being the deuteron channel. 

\begin{table}[b]
\caption{Two-body $\Sigma N$ channels $(i_\Sigma,s_\Sigma)$,
$\Lambda N$ channels $(i_\Lambda,s_\Lambda)$,
$NN$ channels with $\Sigma$ spectator $(i_{N(\Sigma)},s_{N(\Sigma)})$, and
$NN$ channels with $\Lambda$ spectator $(i_{N(\Lambda)},s_{N(\Lambda)})$ that contribute to
a given $J^P=1/2^+$ $\Lambda NN - \Sigma NN$ state with total isospin $I$.}
\begin{ruledtabular}
\begin{tabular}{ccccc}
$I$ & $(i_\Sigma,s_\Sigma)$  & $(i_\Lambda,s_\Lambda)$ 
& $(i_{N(\Sigma)},s_{N(\Sigma)})$ & $(i_{N(\Lambda)},s_{N(\Lambda)})$ \\ 
\hline 
0 & (1/2,0),(1/2,1)  & (1/2,0),(1/2,1)  & (1,0) & (0,1) \\ 
1 & (1/2,0),(3/2,0),(1/2,1),(3/2,1)  & (1/2,0),(1/2,1) & (0,1),(1,0) &
(1,0)  \\ 
2 & (3/2,0),(3/2,1) & & (1,0) &  \\ 
\end{tabular}
\label{t4}
\end{ruledtabular}
\end{table}

\begin{figure}[t]
\vspace*{-0.5cm}
\includegraphics[width=.58\columnwidth]{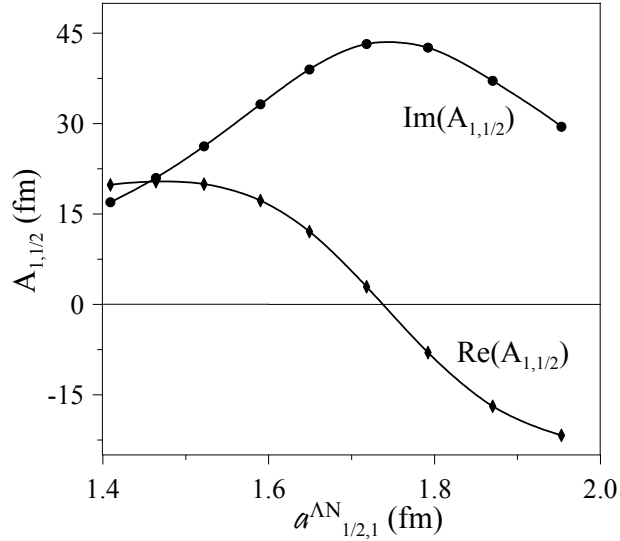}
\vspace*{-6.cm}
\caption{Real and imaginary parts of the $\Sigma d$ scattering
length, $\rm{A_{1,1/2}}$.}
\label{fig2}
\end{figure}

The most interesting result in connection with the results 
reported in~\cite{Pan22} is the  prediction
of a $\Sigma NN$ $(I,J^P)=(1,1/2^+)$ 
quasibound state in the region near threshold. We show in 
Fig.~\ref{fig2} the real, Re$(\rm{A_{1,1/2}})$, and imaginary, Im$(\rm{A_{1,1/2}})$,
parts of the $\Sigma d$ scattering length as a function of the attraction
in the three-body channel. The real part
becomes negative while the imaginary part
has a maximum, which are the typical
signals of a quasibound state~\cite{Del00}.
The position of the pole almost does not change for the 
different models, being at
2.92$\, -i\, $2.17 MeV for the reference model. 
The width of this state comes mainly from the coupling to
a $D$ wave $\Lambda NN$ channel.
It is worth to emphasize that
the enhancement suggested as a possible $\Sigma NN$ 
resonance by the Hall A Collaboration at Jefferson Lab 
appears at about $(3.14 \pm 0.84) - i (2.28 \pm 1.2)$ MeV~\cite{Pan22},
in very good agreement with the theoretical results reported in this study.

The existence of a $(I,J^P)=(1,1/2^+)$ $\Sigma NN$ quasibound state
was suggested in a variational calculation investigating the structure
of the $A=3$ $\Sigma$-hypernuclei~\cite{Koi96}.
Similar results were obtained by Harada and 
Hirabayashi~\cite{Har14} using a distorted-wave impulse approximation 
within a coupled $(2N-\Lambda )+(2N-\Sigma )$ model with a 
spreading potential. 
Afnan and Gibson found a near threshold $I=0$
resonance while exploring $\Lambda d$ elastic scattering by a 
continuum Faddeev calculation~\cite{Afn90}. 
Recent preliminary calculations ~\cite{Gib22} have suggested that
the pole for the $I=1$ resonance is also located near the 
$\Sigma NN$ threshold, but the two resonances are unlikely to be
distinguished experimentally.

To conclude, Hall A Collaboration at Jefferson Lab~\cite{Pan22}
has found indications of the possible existence of
a $\Sigma NN$ resonance at $(3.14 \pm 0.84) - i (2.28 \pm 1.2)$ MeV.
The state is likely a $\Sigma^0 nn$ state, although 
this has to be confirmed by future experiments.
We have presented a detailed study of $\Lambda NN - \Sigma NN$ system
using the hyperon-nucleon and nucleon-nucleon
interactions derived from a chiral constituent quark model with
full inclusion of the $\Lambda \leftrightarrow \Sigma$ conversion
and taking into account all three-body configurations
with $S$ and $D$ wave components.
In the case of the $\Sigma NN$ system there exists a narrow quasibound
state near threshold in the $(I,J^P)=(1,1/2^+)$ channel. 
The position of the pole is at
2.92$\, -i\, $2.17 MeV. There is a reasonable agreement with
the enhancement suggested as a possible $\Sigma NN$ 
resonance by the Hall A Collaboration at Jefferson Lab, 
appearing at about $(3.14 \pm 0.84) - i (2.28 \pm 1.2)$ MeV,
being our result inside the error bar of the experimental data.

\section{Acknowledgments}
This work has been partially funded by COFAA-IPN (M\'exico) and 
by Ministerio de Ciencia e Innovaci\'on and EU FEDER under 
Contracts No. PID2019-105439GB-C22 and RED2018-102572-T.

\end{document}